\documentclass[preprint,pre,showpacs]{revtex4}

\usepackage{amsmath}    
\usepackage{amsfonts}
\usepackage{amssymb}
\usepackage{graphicx}   
\usepackage{dcolumn}
\usepackage{bm}

\begin{document}

\title{Stable transport in proton driven Fast Ignition}

\author{A. Bret}
\email{antoineclaude.bret@uclm.es}
 \affiliation{Instituto de Investigaciones Energ\'{e}ticas y Aplicaciones Industriales, and ETSI Industriales, Universidad de Castilla-La
Mancha, 13071 Ciudad Real, Spain}

\date{\today }

\begin{abstract}
Proton beam transport in the context of proton driven Fast Ignition is usually assumed to be stable due to protons high inertia, but an analytical analysis of the process is still lacking. The stability of a charge and current neutralized proton beam passing through a plasma is therefore conducted here, for typical proton driven Fast Ignition parameters. In the cold regime, two fast growing Buneman-like modes are found, with an inverse growth-rate much smaller than the beam time-of-flight to the target core. The stability issue is thus not so obvious, and Kinetic effects are investigated. One unstable mode is found stabilized by the background plasma protons and electrons temperatures. The second mode is also damped, providing the proton beam thermal spread is larger than $\sim$ 10 keV. In Fusion conditions, the beam propagation should therefore be stable.
\end{abstract}


\maketitle

\emph{Introduction - }
The Fast Ignition Scenario for Inertial Fusion has been proposed more than one decade ago as a solution to relax the symmetry constraints on the target irradiation in conventional schemes \cite{Tabak,Tabak2005}. Such relaxing allows for a lower laser energy, hence higher gain. The first Fast Ignition concept presented implied the propagation of a relativistic electron beam from the border of a pre-compressed target, to the dense core. It was quickly recognized that beam transport stability would be an issue for this scenario \cite{Pukhov1997,califano2}, and many efforts have been devoted to document this point \cite{califano3,Silva2002,HondaPRL,TaguchiPRL}. Proton driven Fast Ignition  was later suggested as a probably more stable alternative, due to the higher inertia of the protons \cite{Roth2001,Roth2009}. The scheme was made possible by the recently discovered possibility of generating intense beams of high energy protons, from the rear surface of solid targets irradiated by ultra-intense lasers \cite{Snavely2000,Hegelich2006,Schwoerer2006,Toncian2006}. This option has been largely investigated since then (see Ref. \cite{Roth2009}, and References therein), and alterative scenarios using more than one beam have already been presented \cite{Temporal2006,Temporal2009}.

Although a proton beam should definitely beam more stable than an electron beam, its stability within the context of proton driven Fast Ignition as not been addressed analytically so far. Further more, some simulation of the transport of a charge and current neutralized proton beam in a dense plasma, suggested that beam-plasma instabilities could also be an issue \cite{Ruhl2004}.
The goal of this paper is to clarify this problem by working out the kinetic theory of the proton beam instability within the alluded scenario. We thus start examining a cold model which shows that the $e$-folding time of the fastest growing mode is indeed much smaller than the beam time-of-flight from the target border to its dense core. In this respect, unstable modes are driven by the interaction of the co-moving electrons with both the beam and the background protons. We thus implement the kinetic theory of the problem accounting for the temperature of the three species, and show that for realistic temperature parameters, the beam propagation should be stable.

\begin{figure}
\begin{center}
 \includegraphics[width=0.45\textwidth]{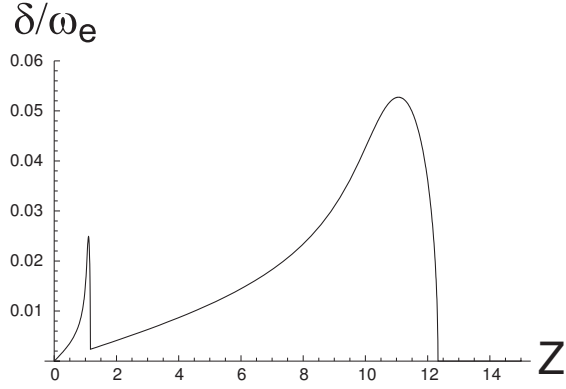}
\end{center}
\caption{growth-rate given by the cold dispersion relation (\ref{eq:dispercold_dimless}) as a function of the reduced wave vector $Z$ for $\alpha=0.1$, and $R=1/1836$.}
\label{fig2}
\end{figure}

\emph{Cold Model - }
We  consider the propagation of a proton beam of density $n_b$ and velocity $v_b$ in a plasma. The plasma proton and electronic densities are $n_p$ and $n_e$ respectively. The beam is initially both current and charge neutralized, so that $n_b+n_p=n_e$ and $n_bv_b=n_ev_e$. Unlike the case of electron driven Fast Ignition, where protons are usually assume at rest due to their higher inertia, the background protons need here to be allowed to move, since the proton beam defines a slower dynamic. Given the typical beam energy of 15 MeV considered in the scenario, protons are not relativistic. A non-relativistic formalism is therefore implemented in the sequel. Such formalism allows us to focus on the flow-aligned instabilities instead of having to search the full 2D unstable spectrum (see discussion in the Conclusion). The cold dispersion equation  for flow-aligned perturbations varying like $\exp (i kz-i\omega t)$ reads \cite{Ichimaru},
\begin{equation}\label{eq:dispercold}
    1=\frac{\omega_p^2}{\omega^2}+\frac{\omega_b^2}{(\omega-kv_b)^2}+\frac{\omega_e^2}{(\omega-kv_e)^2},
\end{equation}
with,
\begin{equation}\label{eq:omega_ps}
    \omega_p^2=\frac{4\pi n_p e^2}{M},~~\omega_b^2=\frac{4\pi n_b e^2}{M},~~\omega_e^2=\frac{4\pi n_e e^2}{m},
\end{equation}
where $M$ and $m$ are the proton and electron masses respectively. We now introduce the dimensionless variables,
\begin{equation}\label{eq:variables}
    x=\frac{\omega}{\omega_e},~~Z=\frac{k v_b}{\omega_e},~~R=\frac{m}{M},~~\alpha=\frac{n_b}{n_p},
\end{equation}
so that the  dispersion equation (\ref{eq:dispercold})  reads,
\begin{equation}\label{eq:dispercold_dimless}
1=\frac{R}{1+\alpha }\frac{1}{x^2}+\frac{R \alpha}{1+\alpha }\frac{1 }{(x-Z)^2 }+\frac{1}{\left(x -Z \frac{\alpha}{1+\alpha} \right)^2}.
\end{equation}
Figure \ref{fig2} displays the growth-rate as a function of $Z$ for $\alpha=0.1$ of the most unstable mode for a given $Z$. The local growth-rate extremum at low $Z$ stems from the interaction of the co-moving electrons with the proton beam. The maximum growth-rate is found here for $Z\sim 1$ and $x\sim 1$ with,
\begin{equation}\label{eq:delta_cold_min}
    \delta=\frac{\sqrt{3}}{2^{4/3}}(R\alpha)^{1/3}\omega_e.
\end{equation}
With $x/Z=\omega/kv_b\sim 1$, this mode is therefore resonant with the beam. The all-spectrum fastest growing mode arises from the Buneman interaction of the electron beam with the background protons \cite{Buneman}. It is located at $Z\sim 1/\alpha$ and $x\sim 0.05$ with,
\begin{equation}\label{eq:delta_cold_max}
    \delta=\frac{\sqrt{3}}{2^{4/3}}R^{1/3}\omega_e.
\end{equation}
With $R=1/1836$, this Buneman mode is found growing at $\delta=0.052\omega_e$. For an initial electronic density $n_e=10^{21}$ cm$^{-3}$, we thus have an $e$-folding time $\tau_e=1/\delta=10^{-14}$ s. This $e$-folding time needs now to be compared with the time the proton beam takes to reach the target core. Assuming the beam needs to travel 100 $\mu$m at $v_b\sim 0.146 c$ (15 MeV), we find a transit time $\tau_t\sim 2\times 10^{-12}$s, so that $\tau_e/\tau_t > 200$.
This simple calculation of the growth-rate shows that the instability could reach 200 $e$-folding times before the beam reaches the target core. We therefore turn to the kinetic theory in order to evaluate how kinetic effects slow down the instability.

\emph{Kinetic Theory - }
In line with the Dawson's picture of the multi-beam model \cite{Dawson}, Eq. (\ref{eq:dispercold}) can be directly adapted replacing the terms corresponding to each species by their kinetic counterpart. For example, the proton beam of density $n_b$ and velocity $v_b$ is replaced by an infinite numbers of ``micro-beams'' of density $n_bf_b(v)dv$ and velocity $v$, where $f_b(v)$ turns to be the beam velocity distribution function. The substitution operated is thus,
\begin{equation}\label{eq:subs}
  \frac{1}{(\omega-kv_b)^2}\rightarrow  \int_{-\infty}^{\infty} \frac{f_b(v)dv}{(\omega-kv)^2},
\end{equation}
with
\begin{equation}\label{eq:Maxwell}
    f_b(v)=\sqrt{\frac{M}{2\pi k_B T_b}}\exp\left[\frac{-M (v-v_b)^2}{2k_B T_b}\right].
\end{equation}
After substitution of the three terms of Eq. (\ref{eq:dispercold}), the kinetic dispersion equation is found as,
\begin{eqnarray}\label{eq:Disper_Kine}
    0=1&+&\frac{2\omega_p^2}{k^2v_{Tp}^2}W\left[\frac{\omega/k}{v_{Tp}}\right]+\frac{2\omega_b^2}{k^2v_{Tb}^2}W\left[\frac{\omega/k-v_b}{v_{Tb}}\right]\nonumber\\
    &+&\frac{2\omega_e^2}{k^2v_{Te}^2}W\left[\frac{\omega/k-v_e}{v_{Te}}\right],
\end{eqnarray}
where,
\begin{equation}\label{eq:W}
    W(\xi)= \frac{1}{\sqrt{\pi}}\int_{-\infty}^{\infty}\frac{t e^{-t^2}dt}{\xi-t}
\end{equation}
and the three thermal velocities read,
\begin{equation}\label{eq:Vths}
    v_{Tp}^2=\frac{2k_BT_p}{M},~~ v_{Tb}^2=\frac{2k_BT_b}{M},~~ v_{Te}^2=\frac{2k_BT_e}{m}.
\end{equation}
Equation (\ref{eq:Disper_Kine}) can be of course equally derived from the linearized Maxwell-Vlasov system of equations. In terms of the dimensionless variables defined in Eqs. (\ref{eq:variables}), the equation reads,
\begin{eqnarray}\label{eq:Disper_Kine_Dim}
    0=1&+&\frac{R}{1+\alpha}\frac{1}{\rho_{Tp}^2Z^2}W\left[\frac{x}{Z\rho_{Tp}}\right]\nonumber\\
    &+&\frac{R\alpha}{1+\alpha}\frac{1}{\rho_{Tb}^2Z^2}W\left[\frac{x/Z-1}{\rho_{Tb}}\right]\nonumber\\
    &+&\frac{1}{\rho_{Te}^2Z^2}W\left[\frac{x(1+\alpha)/Z-\alpha}{\rho_{Te}(1+\alpha)}\right],
\end{eqnarray}
with,
\begin{equation}\label{eq:rho}
    \rho_{Tp}=\frac{v_{Tp}}{v_b},~~ \rho_{Tb}=\frac{v_{Tb}}{v_b},~~ \rho_{Te}=\frac{v_{Te}}{v_b}.
\end{equation}
At this junction, we consider typical proton driven Fast Ignition parameters, namely a quasi mono-energetic proton beam of mean energy 15 MeV, energy spread 1 MeV and density $10^{20}$ cm$^{-3}$, entering a plasma of density $10^{21}$ cm$^{-3}$ and temperature 1 keV for both protons and electrons. These values set the dimensionless parameters defined by Eqs. (\ref{eq:variables},\ref{eq:rho}) to
\begin{equation}\label{eq:rho_val}
    \alpha=0.1,~~\rho_{Tp}=8.10^{-3},~~ \rho_{Tb}=0.25,~~ \rho_{Te}=0.35.
\end{equation}

\begin{figure}
\begin{center}
 \includegraphics[width=0.45\textwidth]{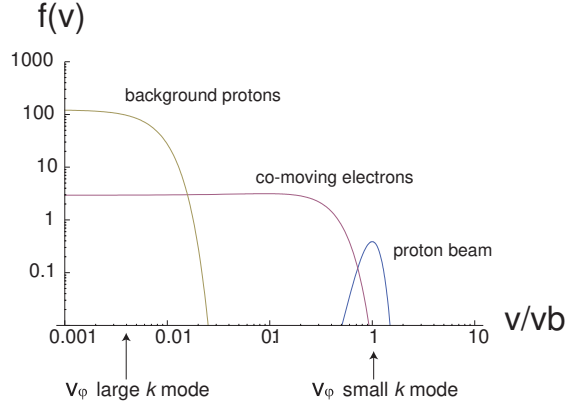}
\end{center}
\caption{(Color online) Distribution functions of the three species involved. The positions of the phase velocities of the two most unstable modes are located.}
\label{fig3}
\end{figure}

The three distribution functions are displayed on Figure \ref{fig3}, allowing for a intuitive evaluation of kinetic effects upon the two aforementioned unstable modes. The large $k$ mode, by virtue of its small phase velocity, is expected to interact mostly with the electrons and the background protons thermal spreads. For this mode, the beam is virtually cold.

As evidenced on the plot, the small $k$ mode is beam resonant, but the large electronic thermal spread should significantly affect it. For this mode, the background protons are almost cold, but the thermal spreads of the remaining species should be important.

\begin{figure}
\begin{center}
 \includegraphics[width=0.45\textwidth]{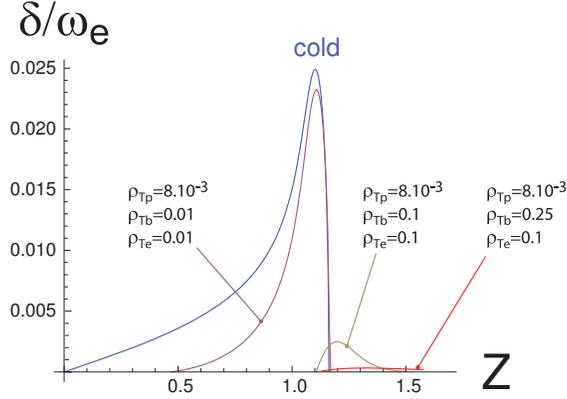}
\end{center}
\caption{(Color online) growth-rate of the low $k$ unstable mode as a function of $Z=kv_b/\omega_p$, and for various temperature parameters.}
\label{fig4}
\end{figure}

\begin{figure}
\begin{center}
 \includegraphics[width=0.45\textwidth]{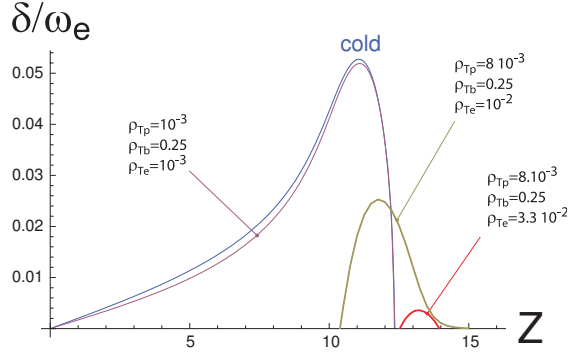}
\end{center}
\caption{(Color online) growth-rate of the large $k$ unstable mode as a function of $Z=kv_b/\omega_p$, and for various temperature parameters.}
\label{fig5}
\end{figure}

These trends are now confirmed by Figures \ref{fig4} and \ref{fig5}, where the numerical evaluation of the growth-rate has been performed for both modes in terms of the reduced wave vector $Z=kv_b/\omega_p$, and for various temperature parameters. From the numerical point of view, the full kinetic dispersion equation (\ref{eq:Disper_Kine_Dim}) has been solved in each case, using the ``FindRoot'' routine of \emph{Mathematica} and providing the cold solution as an initial guess for the algorithm.
Figure \ref{fig4}, documenting the small $k$ growth, has been plotted setting the background protons temperature to its realistic value $\rho_{Tp}=8.10^{-3}$ in every case (except the cold one). As expected, this  mode is thus found virtually insensitive to this thermal effect. But both the beam and the electron thermal spreads contribute here to stabilize the waves.  For a beam temperature parameter $\rho_{Tb}=0.25$, the mode is almost stabilized  with $\rho_{Te}=0.1$, which remains much smaller than the expected value. Complete stabilization is achieved for $\rho_{Te}=0.24$ ($T_e=0.47$ keV), so that the part of the spectrum is completely stabilized by $\rho_{Te}=0.35$ ($T_e=1$ keV). The stabilization of this mode is thus provided by the beam and the electron thermal spread. Setting every temperature parameters to their realistic value, and varying the beam one, it is numerically found that stabilization is achieved from $\rho_{Tb}>2.6\times 10^{-2}$. This means that a beam with a thermal spread smaller than $\sim 10$ keV, far smaller than the 1 MeV observed so far, would not propagate stably.

The evolution of the large $k$ unstable mode is pictured on Figure \ref{fig5}, where the beam temperature parameter is set to its realistic value in any cases, evidencing the small influence it has on overall thermal effects. For these short wavelength Buneman modes, stabilization comes from the background protons and the co-moving electrons thermal spread. For background protons realistic temperature parameters, stabilization is almost achieved at $\rho_{Te}=3.10^{-2}$. Complete stabilization is numerically found from  $\rho_{Te}>3.3\times10^{-2}$ ($T_e=10$ eV), which is far lower than the expected thermal spread. It is important to note that the stabilization of this mode does not rely on the beam quality, and is only provided by the initial plasma temperature.

\emph{Conclusion - }
The stability of an intense proton beam passing through a plasma has been analyzed. In line with previous numerical approaches \cite{Ruhl2004}, the beam is assumed to be both charge and current neutralized. A simple cold model describing the flow aligned unstable modes evidences two Buneman-like unstable modes, arising from the co-moving-electrons/proton-beam, and co-moving-electrons/background-protons, interactions. Noteworthily, the fastest growing mode in the cold regime is found to have an $e$-folding time 2 orders of magnitude \emph{lower} than the beam time-of-flight to the dense core. A kinetic calculation is then undertaken, which shows that for realistic temperature parameters of the three species, flow-aligned instabilities should definitely be stabilized.

As previously mentioned, we focused here on flow-aligned unstable perturbations. When it comes to the unstable modes involved in the case of a relativistic electron beam for example, it is now known that Filamentation or Oblique modes can govern the linear phase \cite{BretPRL2005,frederiksenPoP2008,kong2009}, with a wave-vector not parallel to the beam direction. Such shifting of the dominant mode is mainly due to anisotropic relativistic effects which make the particles ``heavier'' to move along the beam axis than in the perpendicular direction. In the non-relativistic regime examined here, the unstable spectrum is still 2D, but the growth of non-parallel instabilities is hindered by thermal effects which prevent (or simply slow-down) the formation of filamentary structures \cite{Silva2002}. Electrostatic modes growing along the beam direction are thus expected to be the most unstable ones. It could happen that once these modes have been stabilized, the system  remains unstable with respect to the Filamentation instability. Indeed, situations have been evidenced where a relative drift still excites the later, while the formers are stabilized \cite{tzoufras}. At any rate, the growth of Filamentation must be slower in the kinetic regime than in the cold one. The $e$-folding number of the Filamentation instability for the present case has thus been checked, and found lower than 3. This instability should thus not be a threat in the present case.

Finally, we here considered an infinite and homogenous plasmas, while the realistic situation involves a density gradient. In this respect, the uniform plasma approximation is valid providing the wavelengths implied in the instability are short compared to the scale length of the density gradient. This question has been addressed in a previous work  \cite{BretPoPGradient}. On the one hand, the wavelengths relevant to the unstable spectrum are typically the plasma skin depth which is 0.17 $\mu$m at $n=10^{21}$ cm$^{-3}$. On the other hand, an exponential density gradient from the target border to its core sets a gradient scale length $\lambda\sim 10$  $\mu$m. The present approximation is thus valid from the onset of the beam transport, and even more justified at later time since the plasma skin depth is monotonically  decreasing as the plasma density rises.

\emph{Acknowledgements - }
This work has been  achieved under projects FIS 2006-05389 of the
Spanish Ministerio de Educaci\'{o}n y Ciencia and PAI08-0182-3162 of
the Consejer\'{i}a de Educaci\'{o}n y Ciencia de la Junta de
Comunidades de Castilla-La Mancha. Thanks are due to Laurent Gremillet for useful discussions.


\end{document}